# Single track coincidence measurements of fluorescent and plastic nuclear track detectors in therapeutic carbon beams

**Julia-Maria Osinga,**[a,b,*] **Iva Ambrožová,**[c] **Kateřina Pachnerová Brabcová,**[c,d] **Mark S. Akselrod,**[e] **Oliver Jäkel,**[a,b,f] **Marie Davídková,**[c] **and Steffen Greilich**[b]

[a] *Heidelberg University Hospital, Department of Radiation Oncology and Radiation Therapy, Im Neuenheimer Feld 400, 69120 Heidelberg, Germany*

[b] *German Cancer Research Center (dkfz), Division of Medical Physics in Radiation Oncology, Im Neuenheimer Feld 280, 69120 Heidelberg, Germany*

[c] *Nuclear Physics Institute AS CR, Department of Radiation Dosimetry, Na Truhlářce 39/64, 180 00 Prague, Czech Republic*

[d] *Chalmers University of Technology, Department of Applied Physics, SE-41296 Gothenburg, Sweden*

[e] *Landauer Inc., 723 1/2 Eastgate, Stillwater, OK 74074, USA*

[f] *Heidelberg Ion-Beam Therapy Center, Im Neuenheimer Feld 450, 69120 Heidelberg, Germany*

*E-mail*: j.osinga@dkfz.de

ABSTRACT: In this paper we present a method for single track coincidence measurements using two different track detector materials. We employed plastic and fluorescent nuclear track detectors (PNTDs and FNTDs) in the entrance channel of a monoenergetic carbon ion beam covering the therapeutically useful energy range from 80 to 425 MeV/u. About 99 % of all primary particle tracks detected by both detectors were successfully matched, while 1 % of the particles were only detected by the FNTDs because of their superior spatial resolution. We conclude that both PNTDs and FNTDs are suitable for clinical carbon beam dosimetry with a detection efficiency of at least 98.82 % and 99.83 % respectively, if irradiations are performed with low fluence in the entrance channel of the ion beam. The investigated method can be adapted to other nuclear track detectors and offers the possibility to characterize new track detector materials against well-known detectors. Further, by combining two detectors with a restricted working range in the presented way a hybrid-detector system can be created with an extended and optimized working range.

[*] Corresponding author.

## 1. Introduction

Nuclear track detectors are a well-known technology with a wide range of applications [1], such as neutron dosimetry, space dosimetry or fluence measurements in ion beam accelerators. With their ability of energy discrimination, some nuclear track detectors represent also a promising approach for specific tasks in the worldwide growing field of radiotherapy with swift protons and ions [2]. Potential applications are the assessment of biological dose, in-vivo dosimetry as well as employment where the use of ionization chambers is challenging, such as dosimetry in magnetic fields. However, each detector material has certain limitations in efficiency with respect to the linear energy transfer (LET), the atomic number Z or the incident angle of the traversing particle. Thus, for characterization and application of nuclear track detectors some prior knowledge of the beam is needed. By combining two different nuclear track detector materials and especially comparing both detectors on an individual track basis, information regarding the number and characteristics of registered particles can be gained substituting the *a priori* knowledge of beam parameters to a certain extent.

In this paper we, therefore, develop a method for single-track coincidence measurements using the example case of a plastic and fluorescent nuclear track detector (PNTD and FNTD). Irradiations were performed in therapeutic carbon ion beams with the results compared on a track-by-track level (Figure 1). PNTDs are a well-established tool for fluence measurements and LET determination. However, when using optical microscopy rather than more sophisticated methods like atomic force microscopy (AFM), their fluence range is more restricted. Further, the LET range of detectable particles has been considered too narrow for applications in ion beam radiotherapy and their suitability for clinical ion beam dosimetry was, therefore, doubted by some researchers [3-4]. $Al_2O_3$:C,Mg-based FNTDs [5], on the other hand, are expected to cover a LET range from at least $L_\infty(Al_2O_3)$ = 0.5 keV/µm to 61000 keV/µm with a close-to-perfect detection efficiency at particle fluences up to at least $5 \times 10^7$ per $cm^2$ [6]. Thus, FNTDs cover also the detection of lighter particles found as secondary fragments in therapeutic carbon ion beams (*i.e.* case 4 of figure 1 is highly unlikely).

## 2. Materials and methods

### 2.1 Fluorescent nuclear track detectors

We used $Al_2O_3$:C,Mg single crystals grown by Landauer Inc., Stillwater, OK U.S.A. (4×8×0.5 $mm^3$ in size) as FNTDs [5]. $Al_2O_3$:C,Mg crystal contain a high concentration of aggregate $F_2^{2+}$(2Mg) color centers. These centers exhibit radiochromic transformation under ionizing radiation (Figure 2). The resulting $F_2^{+}$(2Mg) centers show intra-center fluorescence at about 750 nm when excited at 620±50 nm with a lifetime of 75±5 ns and a high quantum yield of fluorescence. Further, the transformed centers are optically, thermally, and temporally stable. This enables fast, non-destructive optical imaging of energy deposition by ionizing radiation and thus charged particle tracks with sub-micrometer resolution by means of confocal laser scanning fluorescence microscopy [7]. $F_2^{2+}$(2Mg) color centers also undergo photochromic transformation to $F_2^{+}$(2Mg) by sequential two-photon absorption (2PA) of 435±40 nm light (Figure 2). The transformed centers have the same properties as radiation-induced $F_2^{+}$(2Mg)-centers and can thus be read out in the same way.

### 2.2 Zeiss LSM 710 ConfoCor 3

FNTDs were read-out using the inverted laser scanning confocal microscope Zeiss LSM 710 ConfoCor 3 as described in [8] (633 nm helium-neon laser for excitation, single photon counting avalanche photodiode with 655 nm long-pass emission filter for detection). A lateral (axial) resolution of about 200 nm (800 nm) was obtained with a 63x/1.40NA oil immersion objective lens. Additionally, a 405 nm diode laser was used to write fiducial markers into the FNTD using the photochromic transformation process.

### 2.3 Image processing software

To subtract the fluorescence background of the FNTD images and to determine the particle track positions, we employed the 'Mosaic' background subtractor [9] and particle tracker plug-in [10] for the software ImageJ [11], [12] as described in [6]. Further data processing, image registration, and track matching was done in R (version 2.14.2) [13] with the 'FNTD' extension package.

### 2.4 Plastic nuclear track detectors

As PNTDs we used HARZLAS TD-1 (poly(allyl) diglycol carbonate, $20\times10\times0.9$ mm$^3$ in size) manufactured by Fukuvi Chemical Industry, Japan. Heavy ionizing particles create microscopic damage (a 'latent track') in the detector material, which can be enlarged by chemical etching and observed under an optical microscope [14].

Within this study, the detectors were etched after irradiation in 5M NaOH at 70$^o$C for 18 hours, which corresponds to the removal of a layer of about 15.5 µm on each side of the PNTD. Under these etching conditions, the optimal fluence range of this PNTD is in the order of $10^3$–$10^5$ cm$^{-2}$. Etched detectors were processed with the high speed microscope system HSP-1000 and the HspFit software [15] providing information about each track´s coordinates, dimension (major and minor axis, diameter, and area), shade *etc*.

### 2.5 Coincidence measurements

Figure 3 shows schematically the workflow of the coincidence measurements. To register the results from both detectors, a relative coordinate system was established prior to irradiation. Three fiducial markers were applied to each detector, optically for the FNTDs and mechanically for the PNTDs. The relative position of the detectors was fixed by taping the detectors back-to-back. The readout sides of both detectors were facing each other in order to minimize a subsequent displacement of the tracked particle positions. After irradiation, both detectors were separately read out and further processed. A transformation matrix covering translation and rotation was derived using the fiducial positions (Figure 4a) to handle the particle position from both detectors in the same coordinate system.

### 2.6 Track-level matching

A critical distance $d_{crit}$ was defined as the maximum Euclidean distance between two tracks on the detectors to be considered as matches. Because uncertainties of the fiducial positions used for image registration give rise to a residual offset between the detectors, boundary effects were taken into account (Figure 4b). The residual offset was additionally minimized using optimization of the number of tracks matched (see supplementary information for details).

### 2.7 Data visualization

Detector images were overlaid with the matching results, with different colors and shapes indicating the cases described in figure 1 (Figure 5a-c). Areas with non-matches were automatically selected for reviewing.

## 3. Experiments

### 3.1 Irradiation

Seven detector pairs were irradiated with carbon ions ($^{12}$C) at the Heidelberg Ion-Beam Therapy Center (HIT). The detector packages were located in the iso-center of the beam with the FNTD facing the beamline. The particles impinged perpendicularly onto the detectors to minimize effects due to the angular dependency of the detection efficiency of PNTDs. For this first study, all irradiations were performed under ideal conditions, *i.e.* perpendicular, without ripple filter in the entrance channel of the carbon ion beam for mono-energetic fields at a depth of 4.54 mm water-equivalent thickness (WET) (2.89 mm WET for the beam application system including monitor chambers and beam exit window plus 1.65 mm WET for the FNTD). The field size was chosen to $10\times10$ cm$^2$. Particle energy and fluence allowed for optimal operation of the PNTD with the procedure used within this study for most measurements (Table 1) except for two irradiations (detector pair 6 and 7) at nominal fluences of $10^6$ and $10^7$ $cm^{-2}$. As the detection efficiency of FNTDs has been determined to > 99.83 % in case of 20 MeV protons [6], it is reasonable to assume a similar performance concerning the irradiations performed during this study due to the higher LET range.

### 3.2 Detector readout and particle tracking

**FNTD**

All images were acquired approximately 30 µm below the sample surface (corresponding to 100 µm WET). In case of the detectors irradiated with low fluence ($10^5$ $cm^{-2}$), a z-stack of three tiled images separated by $\Delta z = 3$ µm and covering an area of 1.26 $mm^2$ was evaluated (Figure 6). The area comprised more than 1000 particle tracks and thus allows to study per mill effects [6]. Thus, although only a limited number of irradiations (7) have been performed, this high number of particle tracks on each detector per irradiation offers good statistics. Where applicable, a maximum intensity projection of the images obtained in depth was produced to further enlarge the signal-to-noise ratio [6]. After background subtraction, the 'Mosaic' particle tracker was applied to the images and corrected manually for

- '**tile scan artifacts**', related to mechanical inaccuracies of the microscope (for details see figure 5a),

and

- **FNTD tracking failures,** *i.e.* either tracks that were actually detected but missed by the automatic particle tracker (false negative), or 'imaginary' tracks counted by mistake (false positive).

For the low fluence irradiations these effects applied in average to 1.68 % of all tracks found on the FNTDs ('tile scan artifacts' 1.60 %; FNTD particle tracker failures: false positive 0.04 %, false negative 0.04 %). Due to the increasing impact of 'tile scan artifacts' with an increased fluence, multiple (4-6) single images (225×225 $\mu m^2$ each) instead of one tiled image were acquired for the detectors irradiated with high fluence ($10^6$ and $10^7 cm^{-2}$). The settings for the z-stack remained the same.

**PNTD**

After the etching procedure, an area of about 2.0×5.5 $mm^2$ was evaluated. All etched particle tracks were analyzed automatically in a first step and afterwards corrected manually for false positive and false negative tracking failures where necessary. Only the detector pairs 1–5 were analyzed; detectors irradiated at higher fluences ($10^6$ and $10^7$ $cm^{-2}$) were impossible to evaluate due to overlapping of tracks.

### 3.3 Image registration and matching

PNTD track positions were transformed to the FNTD coordinate system and $d_{crit} = 15$ µm, $d_{sec} =$ 15 µm (see supplementary information for details) was used for matching. For the PNTD fiducials we assumed a position uncertainty of ±20 µm, while we neglected the corresponding uncertainties for the FNTD position (< 1 µm).

## 4. Results

### 4.1 Low fluence

The detector pairs 1-5 could be matched with a mean accuracy of better than 3 μm given by the overall mean Euclidean distance between matched particle tracks. In summary, 98.95 % of the primary particle tracks were successfully matched. 1.01 % were only detected by the FNTDs, while a small percentile of 0.04 % was only detected by the PNTDs (Table 2). No obvious dependence on the beam energy was observed.

All unmatched particle tracks were reviewed manually, which allowed us to identify two cases in which tracks were ineligible for coincidence analysis:

- **Blobs (case 6 in figure 1):** atypically big tracks only detected by the PNTD. Since this kind of structure is seen on unirradiated PNTDs as well, we believe those blobs are false positive tracks due to internal imperfections of the PNTD material (Figure 5b).
- **Fragments (case 3, 5, 7 in figure 1)**: atypically small and less intense tracks only detected by the FNTD. An analysis of the particles' trajectories in the crystal volume has revealed that most of those tracks propagate at an angle with respect to the primary beam direction and are therefore believed to belong most likely to lighter fragments, which have been created within the FNTD. There are three possible reasons why a corresponding particle track has not been found on the PNTD: (i) Since some particles did not traverse the FNTD in the direction of the primary beam (about 56 % of the particles classified as fragments), they might actually have been detected by the PNTD but outside of the area analyzed (Figure 1, case 7). (ii) The LET of the fragments was below the LET threshold of the PNTD (Figure 1, case 3). (iii) The fragments already stopped within the material of the FNTD (Figure 1, case 5).

Additionally, we identified three cases for the non-matching of tracks:

- **Insufficient spatial resolution of PNTD**s: Since tracks are much bigger on the PNTD as compared to the FNTD, closely spaced particle tracks show major overlap and cannot be resolved (Figure 5c). Contrary, FNTDs can resolve adjacent particle tracks on a much smaller scale.
- **Irregular tracks on PNTD:** tracks are atypically shaped and less intense on the PNTD, while the corresponding particle tracks on the FNTD are regularly formed. We assume that this effect might be due to the non-uniformity of PNTD material or irregularities in the etching procedure.
- **Tracks missed by FNTD/PNTD due to unknown reason.**

Figure 7 summarizes the abundance of these effects for the low fluence irradiations. The main reason causing the PNTDs to miss particle tracks is their insufficient ability to discriminate neighbouring particles whose etched tracks show significant overlap. We believe it is safe to exclude the possibility that the particles missed by the PNTDs are due to transversal under a critical angle. In this case the FNTD would have shown a distinct, ellipsoid signature [19]. Nevertheless, even this effect amounts only to 0.88 % with respect to the total number of tracks eligible for matching. All other investigated effects have shown to be technically negligible.

### 4.2 High fluence

For fluences $\Phi \geq 10^6$ cm$^{-2}$ PNTDs show major particle track overlap, which made particle tracking infeasible with the etching procedure used (Figure 5d-e). In contrast, FNTDs clearly resolve single particle tracks at these high fluences corresponding to practical therapeutic doses.

## 5. Discussion

The results show that the suggested method for single track coincidence measurements demonstrated using the example of FNTDs and PNTDs allows for reliable and accurate matching of individual particle tracks. Unambiguous track matching could be performed with a positioning accuracy better than 3 µm and the random track pattern additionally serving as distinct "fingerprint". By automatically analysing over 1000 particle tracks per detector pair, the developed method allows to study per mill effects.

With the detection efficiency of FNTDs being close to 100 % [6] we found a detection efficiency of about 99 % for the PNTDs for these conditions. The data presented here do thus not support the findings of Fukumura et al. [3], namely that the accuracy of PNTDs in the entrance channel of a clinical carbon beam is hampered by incomplete detection of low-LET particles such as fragments. These would have been registered by the FNTDs. This discrepancy between fluence-based and ionization-based dosimetry was recently investigated and discussed in a more detailed study [20].

## 6. Conclusion

It has been shown that the presented method for single track coincidence measurements between two or multiple nuclear track detectors enables to gather a variety of information about a detector material such as detection efficiency, spectroscopic properties, agreement of track positions, *etc.* on an individual track level. The adaption of the investigated method to other nuclear track detectors offers the possibility to characterize new track detector materials against well-known detectors. Further, by combining two detectors with a restricted working range in the presented way a hybrid-detector system can be created with an extended and optimized working range. This could be beneficial for example for accurate track-based dosimetry and beam characterization of clinical proton and ion beams at therapeutic depths.

The exemplary track-by-track comparison of PNTDs and FNTDs has shown that PNTDs are suitable for clinical carbon beam dosimetry with a detection efficiency of about 99 %, if measurements are performed (1) in the entrance channel of the ion beam (small percentile of low-LET fragments) and (2) low fluences are used (less than $10^5$ per cm$^2$ for the PNTDs in this study).

## Acknowledgments

The authors would like to thank Dr. Stephan Brons from HIT for his help with the irradiations as well as Dr. Felix Bestvater and Manuela Brom from the DKFZ light microscopy facility for their long-standing support of the FNTD project.

## References

[1] R.L. Fleischer, P.B. Price and R.M. Walker, *Nuclear tracks in solids: principles and applications*, Universtity of California Press, Berkeley 1975.

[2] Particle Therapy Co-operative Group. PTCOG Home [online], http://ptcog.web.psi.ch (2013).

[3] A. Fukumura et al., *Carbon beam dosimetry intercomparison at HIMAC*, *Phys. Med. Biol.* **43** (1998) 3459-63.

[4] G.H. Hartmann et al., *Results of a small scale dosimetry comparison with carbon-12 ions at GSI Darmstadt*, Proc. Int. Week on Hadrontherapy and 2nd Int. Symp. on Hadrontherapy, Elsevier, (1997) 346-50.

[5] M.S. Akselrod and G.J. Sykora, *Fluorescent nuclear track detector technology - A new way to do passive solid state dosimeter, Radiat. Meas.* **46** (2011) 1671-9.

[6] J.M. Osinga et al., *High-accuracy fluence determination in ion beams using fluorescent nuclear track detectors, Radiat. Meas.* **56** (2013) 294-298.

[7] M.S. Akselrod, A.E. Akselrod, S.S. Orlov, S. Sanyal S and T.H. Underwood, *Fluorescent aluminum oxide crystals for volumetric optical data storage and imaging applications, Journal of Fluorescence* **13(6)** (2003) 503-11.

[8] S. Greilich et al., *Fluorescent nuclear track detectors as a tool for ion-beam therapy research, Radiat. Meas.* **56** (2013) 267-272.

[9] J. Cardinale, Histogram-based background subtractor for ImageJ, ETH Zurich, Switzerland (2010).

[10] I.J. Sbalzarini and P. Koumoutsakos, *Feature point tracking and trajectory analysis for video imaging in cell biology, Journal of Structural Biology* **151** (2005) 182-95.

[11] W.S. Rasband, ImageJ (version 1.46a), U.S. National Institutes of Health, Bethesda, Maryland, U.S.A., http://rsbweb.nih.gov/ij/ (1997-2011).

[12] M.D. Abràmoff, P.J. Magalhães and S.J. Ram, *Image processing with ImageJ, Biophotonics International* **11** (2004) 36-42.

[13] R Development Core Team. R: A language and environment for statistical computing. R Foundation for Statistical Computing, Vienna, http://www.R-project.org (2010).

[14] S.A. Durrani and R.K. Bull, *Solid state nuclear track detection – principles, methods and applications*, Pergamon press, Oxford, New York 1987.

[15] N. Yasuda et al., *Development of a high speed imaging microscope and new software for nuclear track detector analysis, Radiat. Meas.* **40** (2005) 311-5.

[16] S. Greilich, L. Grzanka, N. Bassler, C.E. Andersen and O. Jäkel, *Amorphous track models: a numerical comparison study, Radiat. Meas.* **45** (2010) 1406-9.

[17] Stopping powers and ranges for protons and alpha particles. ICRU Report No. 49. Bethesda: International commission on Radiation Units and Measurements (1993).

[18] Stopping of ions heavier than helium. ICRU Report No. 73. Bethesda: International commission on Radiation Units and Measurements (2005).

[19] M. Niklas, J.A. Bartz, M.S. Akselrod, A. Abdollahi, O. Jäkel and S. Greilich, Ion track reconstruction in 3D using alumina-based fluorescent nuclear track detectors, *Phys. Med. Biol.* **58** (2013) N251-266.

[20] J.M. Osinga, S. Brons, J.A. Bartz, M.S. Akselrod, O. Jäkel and S. Greilich, *Absorbed dose in ion beams: comparison of ionization- and fluence-based measurements*, arXiv:1306.1552v2 [physics.med-ph] (2013).

## 7. Supplementary information

### Track-level matching routine

For automatized matching of corresponding track position from the two detectors $det_1$ (here: FNTD) and $det_2$ (here: PNTD), the following steps were conducted:

1. A critical distance $d_{crit}$ is set, defining the maximum euclidean distance between track positions to be considered as matches.
2. Tracks located close to the boundary of the analyzed area might not be matched, if the corresponding particle track on the other detector is located slightly outside the analyzed area (Figure 4b). Therefore, the analyzed area $A_{total}$ of $det_1$ is reduced by a security distance $d_{sec}$ (Figure 8), yielding the reduced area $A_{red} = A_{total}$-$A_{sec}$.
3. Three datasets are generated:
   - ds.1: Particle track positions of $det_1$ located in $A_{red}$.
   - ds.2: Particle track positions of $det_1$ located in $A_{sec}$.
   - ds.3: Particle track positions of $det_2$ located in $A_{total}$.
4. Find matches

For each particle track in ds.1 the euclidean distance to all particle tracks in ds.3 is calculated. The particle track in ds.3 with the minimal distance is defined as match, if the distance is smaller than the critical distance defined before.

5. Find duplicates

Step 4 does not exclude two particles tracks in ds.1 having the same matching particle track in ds.3. Therefore, the matches with the minimal distance are considered to be the correct ones while all other positions becoming unmatched again.

6. Step 4 and 5 are repeatedly applied to ds.1 and the unmatched particle tracks in ds.3 until no more duplicates are found.
7. Due to the residual offset between $det_1$ and $det_2$ it is possible, that remaining unmatched particle tracks in ds.3 have a matching particle track in $A_{sec}$ of $det_1$, which has not been considered during the previous steps. Therefore, step 4 - 6 are applied again considering only ds.2 and the remaining unmatched particle tracks in ds.3.

The previous described matching approaches minimize the boundary effect caused by the residual offset between $det_1$ and $det_2$, but do not reduce the offset itself. In order to do so, an optimization routine was developed enveloping the image registration process as well as the matching steps 1 - 7. In an overall iteration process, the positions of the markers are corrected within the given uncertainties. As a result, the calculated transformation matrix changes as well, which has a direct influence on the matching efficiency. The positions of the markers are optimized that way, that the number of matched particle tracks increases. Consequently, the residual shift between $det_1$ and $det_2$ decreases being beneficial for the overall matching efficiency. The optimal cross positions are found when the number of matched particle tracks is maximal.

## Figures

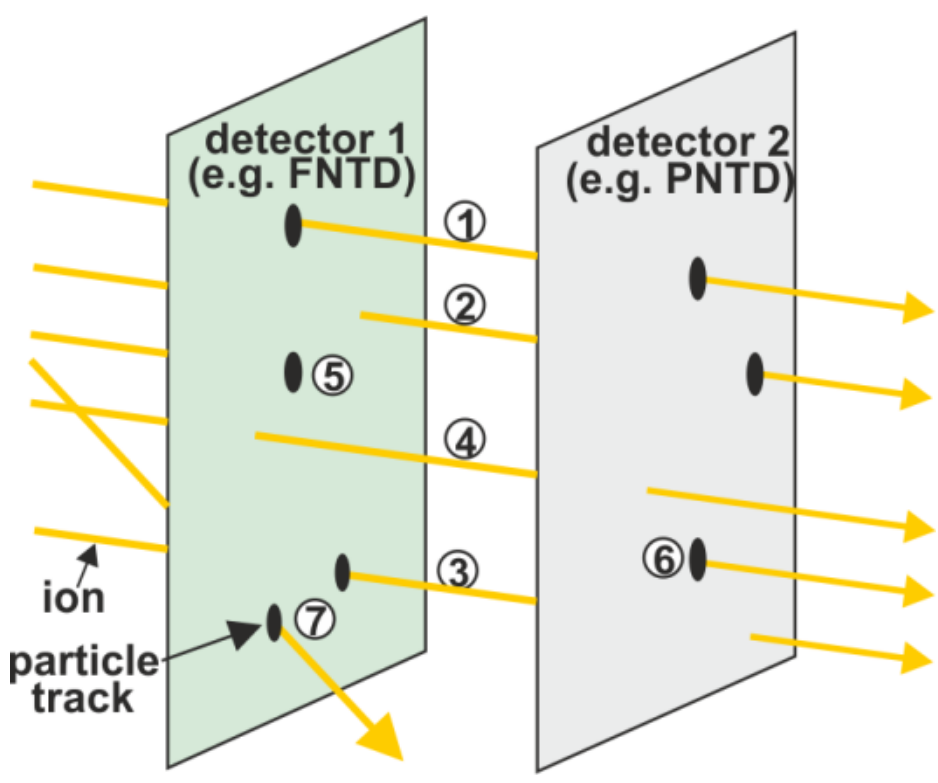

**Figure 1:** Synchronous measurement using two track detectors. Coincident events: ***Case 1:*** Both detectors register a particle. ***Case 2,3:*** Only one of the detectors registers a particle. ***Case 4:*** A particle is not registered by either detector. Non-coincident events: ***Case 5:*** Particle stopping in detector 1. ***Case 6:*** Particle starting in detector 2. ***Case 7:*** Large polar angle, particle misses area considered on one of the detectors.

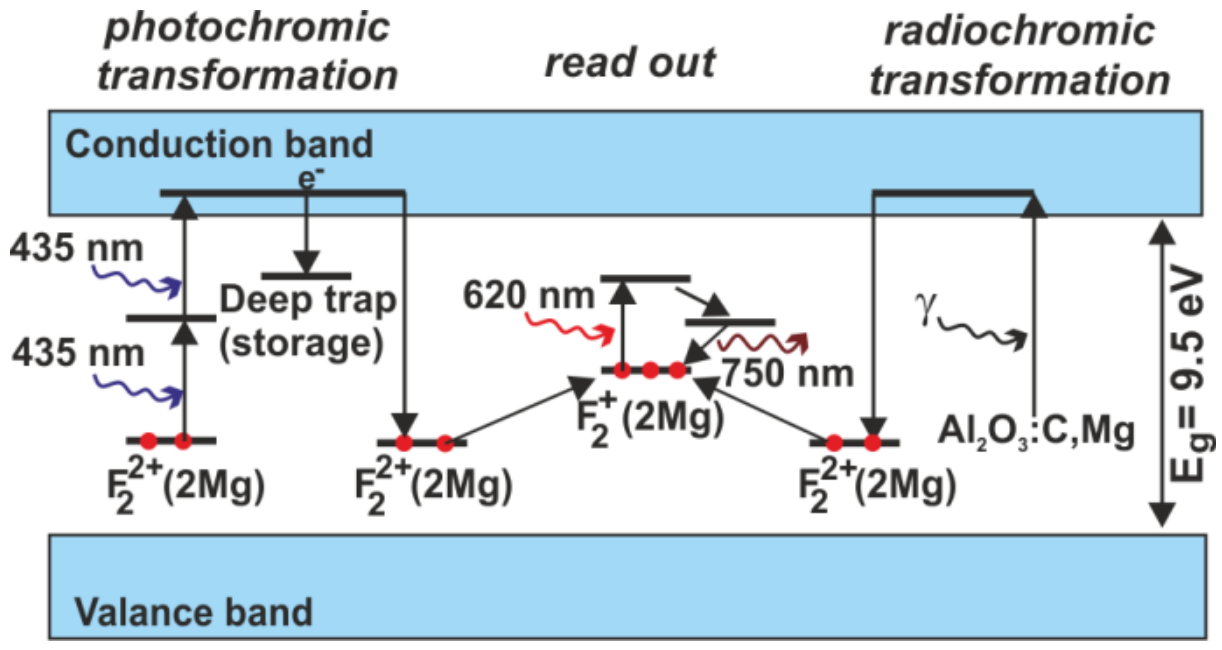

**Figure 2:** Band diagram and electronic processes in $Al_2O_3$:C,Mg used within this study. ***Radiochromic transformation:*** Under ionizing radiation $F_2^{2+}$(2Mg)-center capture an electron and become $F_2^{+}$(2Mg)-center. ***Photochromic transformation:*** The first absorbed photon transfers an electron of the $F_2^{2+}$(2Mg)-center to its excited state. This state is metastable having a lifetime of 9 ns and thus allowing for sequential two photon absorption (2PA). A second photon of the same wavelength arriving within this lifetime window performs a photo-ionization of the center by inducing the second transition between the excited state and the conduction band. The released electron is then captured by a deep trap allowing for long-term storage, which is most probably formed by a similar $F_2^{2+}$(2Mg)-center causing a photochromic transformation and thus the creation of a new three-electron $F_2^{+}$(2Mg)-center. ***Detector read out:*** The transformed $F_2^{+}$(2Mg)-center can be excited with red laser light resulting in a localized transition followed by a radiative decay producing near-infrared fluorescence. Simplified reproduction according to [7].

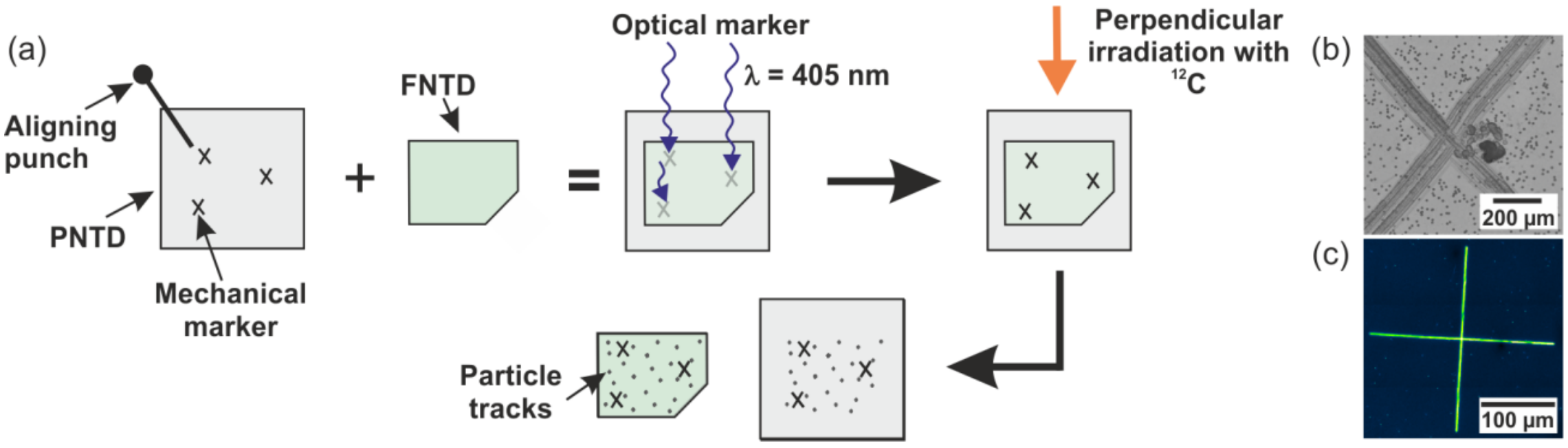

**Figure 3:** (a) Scheme of the overall workflow for the coincidence measurements between FNTDs and PNTDs. First, fiducials were scratched into the PNTD, which become clearly visible after etching the detector (b). Then, an FNTD was put on top of the marked PNTD. After locating the fiducials on the PNTD with the microscope, the FNTD was marked at the corresponding locations using the 405 nm laser of the confocal microscope for the writing process (c).

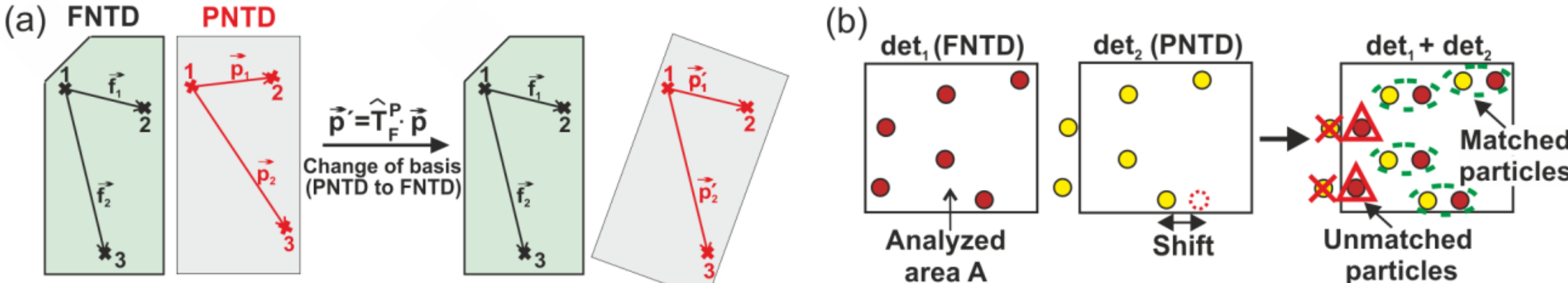

**Figure 4: (a)** Change of basis from the PNTD system into the FNTD system. **(b)** Consequences of a residual shift between the two detectors on the matching efficiency. Particle tracks located close to the boundary of the analyzed area are not matched, if the corresponding particle track on the other detector is located slightly outside the analyzed area.

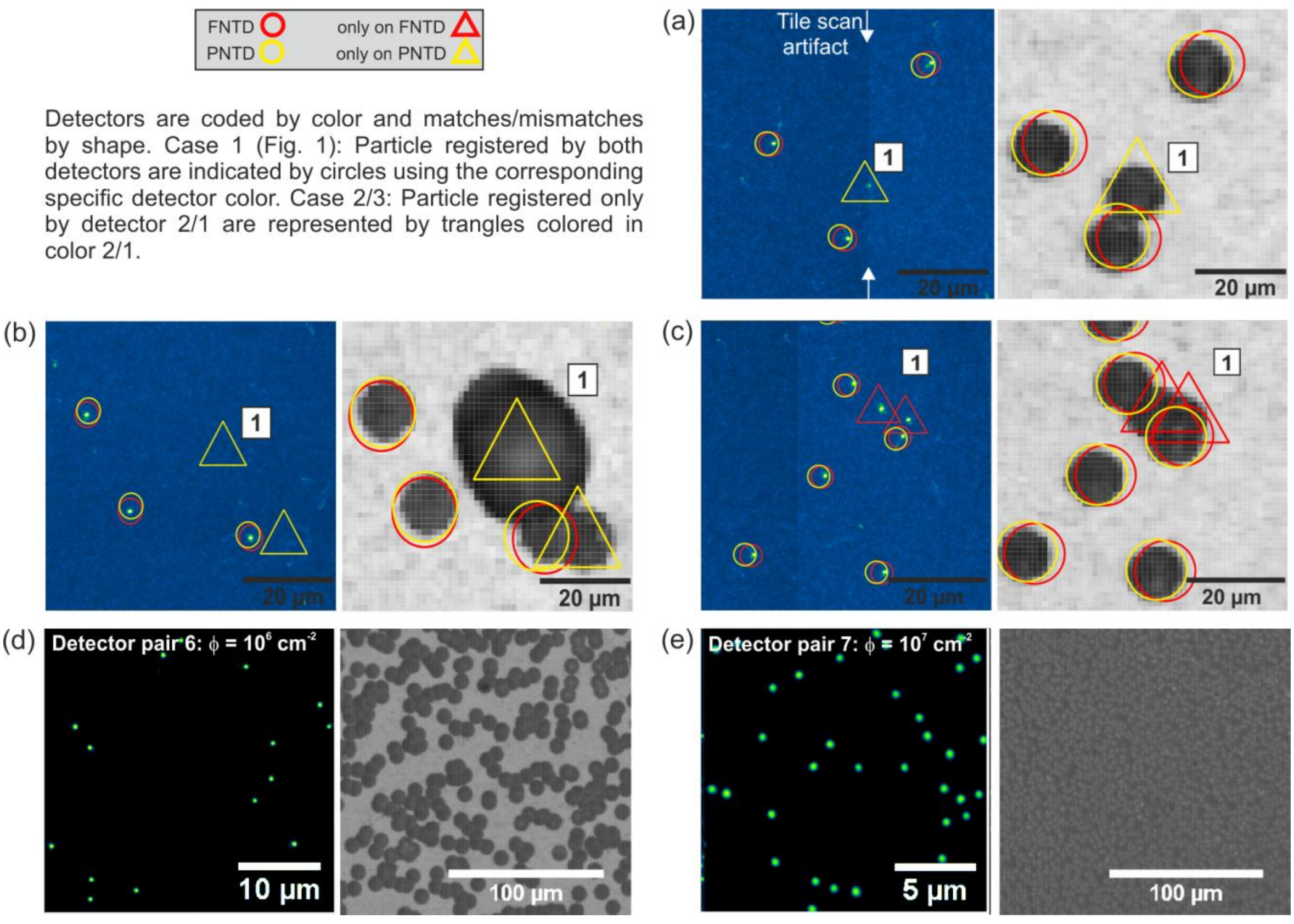

**Figure 5:** Original detector images (right: PNTD, left: FNTD) **(a)**-**(e)** overlaid with the corresponding matching results **(a)**-**(c)**. **(a)** A 'tile scan artifact' is shown on the FNTD (1). Those artifacts are caused by the microscope overlapping adjacent frames of a tiled image (Figure 6) by a few pixels. Because of this overlap, some particle tracks, although successfully detected by the FNTD, get (partly) lost between neighboring frames and are therefore missed by the 'Mosaic' particle tracker. (b) A big, dark 'blob' is shown on the PNTD (1), which has been counted by the PNTD particle tracking routine. Contrary, no particle track is visible on the FNTD. **(c)** The superior resolution of the FNTD (1) is emphasized. **(d)**-**(e)** Images of the detector pairs 6 and 7 irradiated with a fluence of $10^6$ $cm^{-2}$ and $10^7$ $cm^{-2}$ , respectively.

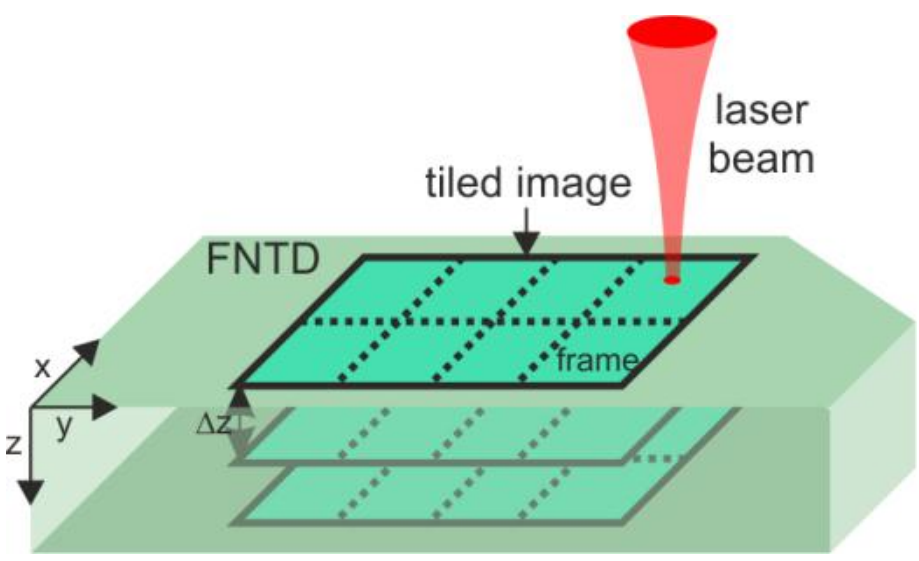

**Figure 6:** Cartoon of the FNTD image acquisition procedure. Tiled images consisting of multiple frames were obtained to cover larger areas. In addition, three layers in depth (separated by Δz) were acquired (referred to as 'z-stack'). Application of a maximum intensity projection for these three layers increases the signal-to-noise ratio.

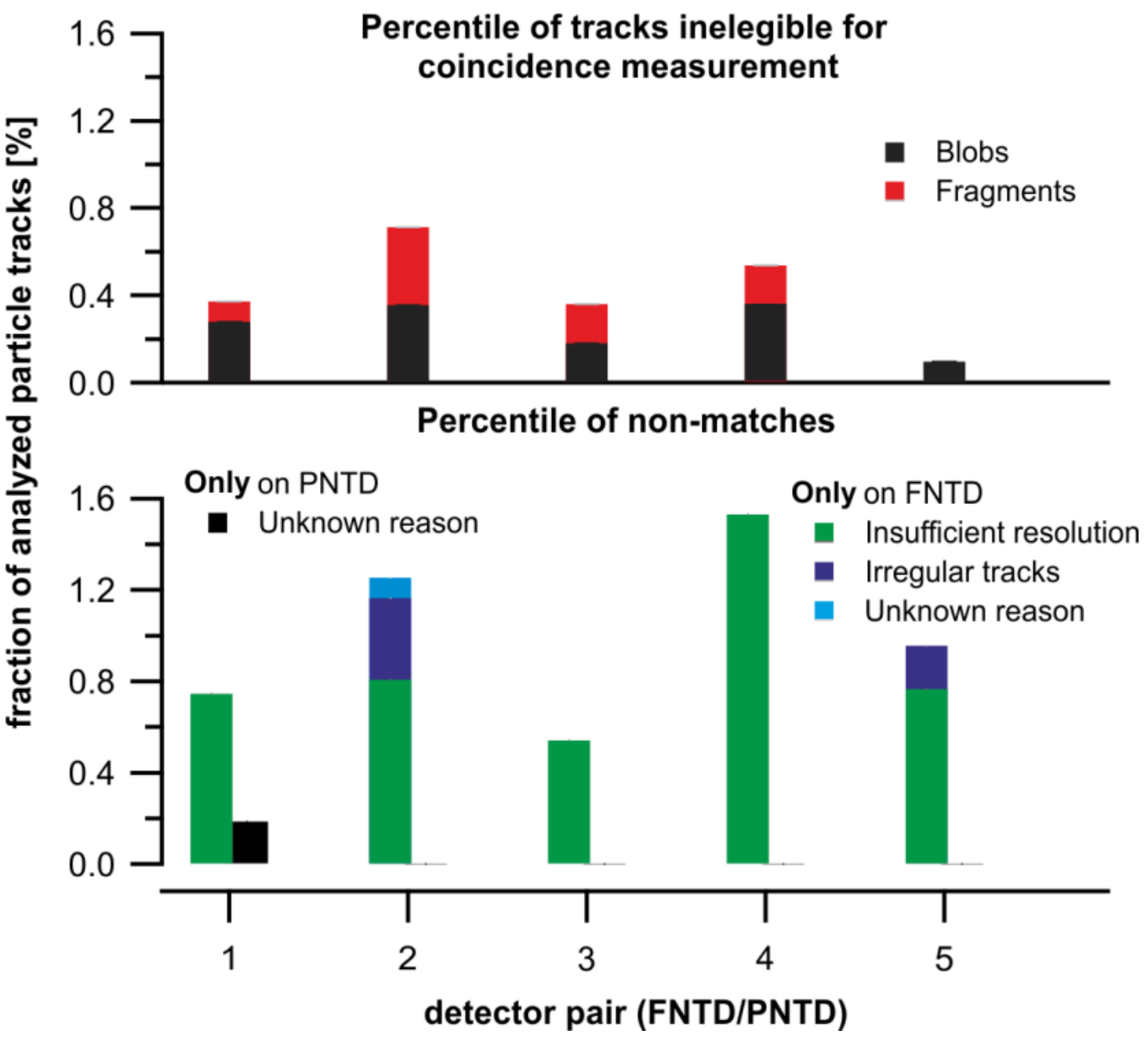

**Figure 7:** Summarized matching results for the detector pairs 1-5 irradiated with low fluence.

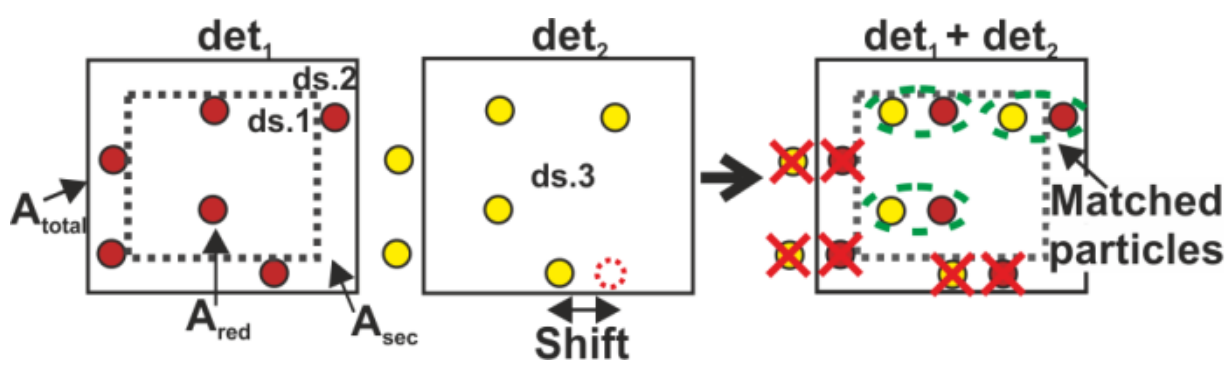

**Figure 8:** Schematic illustration of the tools used to correct for residual boundary effects.

# Tables

| Detector pair | $E_{prim}$(CSDA) [MeV/u] | Fluence [1/cm$^2$] | $s_w/\rho_w$ [keV/µm] |
|---|---|---|---|
| **1** | 80.05 | $10^5$ | 31.37 |
| **2** | 105.40 | $10^5$ | 25.81 |
| **3** | 144.62 | $10^5$ | 20.55 |
| **4** | 219.11 | $10^5$ | 15.56 |
| **5** | 424.78 | $10^5$ | 10.81 |
| **6** | 80.05 | $10^6$ | 31.37 |
| **7** | 80.05 | $10^7$ | 31.37 |

**Table 1:** Overview of the performed $^{12}$C irradiations. Following water equivalent thicknesses (WET) were considered for the calculation of the particle energy at the detector interface ($E_{prim}$) using the continuous slowing down approximation (CSDA) by the "libamtrack" library [16]: (1) 2.89 mm WET, which includes all traversed materials between the high energy beam line and the iso-center, (2) 1.65 mm WET for the FNTD. The mass stopping power values of water were taken from the ICRU reports 49 and 73 [17], [18].

| Detector pair | Total tracks analyzed | Tracks ineligible for coincidence analysis | Eligible tracks for coincidence analysis | Matched tracks | Tracks only on PNTD | Tracks only on FNTD |
|---|---|---|---|---|---|---|
| **1** | 1079 | 4 (0.37 %) | 1075 | 1065 (99.07 %) | 2 (0.19 %) | 8 (0.74 %) |
| **2** | 1124 | 8 (0.71 %) | 1116 | 1102 (98.75 %) | 0 (0.00 %) | 14 (1.25 %) |
| **3** | 1111 | 4 (0.36 %) | 1107 | 1101 (99.46 %) | 0 (0.00 %) | 6 (0.54 %) |
| **4** | 1116 | 6 (0.54 %) | 1110 | 1093 (98.47 %) | 0 (0.00 %) | 17 (1.53 %) |
| **5** | 1046 | 1 (0.10 %) | 1045 | 1035 (99.04 %) | 0 (0.00 %) | 10 (0.96 %) |
| **Average** | **1095** | **5 (0.46 %)** | **1091** | **1079 (98.95 %)** | **0 (0.04 %)** | **11 (1.01 %)** |

**Table 2:** Summarized matching results of the track-by-track comparison study between PNTDs and FNTDs. Total tracks analyzed correspond to all tracks found on the detectors without 'tile scan artifacts' and 'tracking failures'.